\documentclass[a4paper,twocolumn,superscriptaddress,groupedaddress]{revtex4}
\usepackage{amsmath}
\usepackage[hidelinks,urlcolor=blue]{hyperref}
\usepackage{notoccite}

\usepackage{color}

\definecolor{Gray}{gray}{.25}

\usepackage{graphicx}
\graphicspath{.}

\usepackage{sidecap}

\usepackage{wrapfig}
\usepackage[pscoord]{eso-pic}
\usepackage[fulladjust]{marginnote}
\reversemarginpar

\begin{document}

\title{Machine Learning for the edge energies of high symmetry Au nanoparticles}

\author{Emmanouil Pervolarakis}
\email{emper@materials.uoc.gr}
\address{Department of Materials Science and Technology, University of Crete, Heraklion 70013, Greece}

\author{Georgios A. Tritsaris}
\affiliation{John A. Paulson School of Engineering and Applied Sciences, Harvard University, Cambridge, Massachusetts 02138, United States of America}

\author{Phoebus Rosakis}
\address{Department of Mathematics and Applied Mathematics, University of Crete, Heraklion 70013, Greece}
\address{Institute of Applied \& Computational Mathematics, Foundation  for Research and Technology Hellas, 71110 Heraklion, Crete, Greece}

\author{Ioannis N. Remediakis}
\address{
Department of Materials Science and Technology, University of Crete, Heraklion 70013, Greece}
\address{Institute of Electronic Structure and Laser, Foundation for Research and Technology Hellas, 71110 Heraklion, Crete, Greece}

\begin{abstract}
We present data-driven simulations for gold nanostructures, and develop a model that links total energy to geometrical features of the particle, with the ultimate goal of deriving reliable edge energies of gold. Assuming that the total energy can be decomposed into contributions from the bulk, surfaces, edges, and vertices, we use machine learning for reliable multi-variant fits of the associated coefficients. The proposed method of total energy calculations using machine learning produces almost ab-initio-like accuracy with minimal computational cost. Furthermore, a clear definition and metric for edge energy is introduced for edge-energy density calculations that avoid the troublesome definition of edge length in nanostructures. Our results for edge-energy density are 0.22 eV/\AA\ for (100)/(100) edges and 0.20 eV/\AA\  for (111)/(111) edges. Calculated vertex energies are about 1 eV/atom. The present method can be readily extended to other metals and edge orientations as well as arbitrary nanoparticle shapes. 
\end{abstract}

\maketitle


\section{Introduction}

Metallic nanoparticles, in particular those of Au, are key components of modern nanotechnology, mainly owing to their catalytic and plasmonic properties\cite{Atwater2010,Kim2019,Gelle2020}. As it is the case with most materials at the nanoscale, these properties have strong size dependence, since the nanoparticle size governs its volume and surface-to-volume ratio, which in turn determine most physical and chemical properties. In addition to size-dependence, recent studies have shown strong shape-dependence of both chemical and physical properties \cite{Burda2005, Tritsaris2011, Barmparis2016, Li2017}. For example, different shapes exhibit different Localized Surface Plasmon Resonance (LSPR) \cite{Yu2017} and have different numbers of active sites for catalysis since different facets are exposed \cite{Campbell2002}.   Shape-controlled synthesis is usually achieved by fine-tuning the reaction conditions while using suitable ligands \cite{Grzelczak2008, Xia2009}.

Understanding the factors that determine nanoparticle shapes, including the types and lengths of their edges, is crucial for the design of novel nanomaterials.  Shape is known to affect many applications like catalysis, since for different shapes differently coordinated atoms are exposed and different coordinations may be more efficient for certain reactions. Examples where nanoparticle edges play important role include conversion of exhaust gases, sensing and CO$_2$ reduction, just to name a few. Catalytic converters in cars contain metal nanoparticles that catalyse the conversion of poisonous exhaust gases, such as CO, to non-toxic gases, such as CO$_2$. This conversion takes place solely on edges of the nanopartices \cite{lopez04,roling18}. CO$_2$ can be transformed to useful chemicals, including fuels, by means of catalytic CO$_2$ reduction reaction; such reactions take place mostly on edges of nanoparticles \cite{bagger17,bagger19}. This reaction requires a considerable amount of energy, and small Au clusters and nanowires, where the percentage of edge atoms is very high, have shown a lot of promise in reducing that barrier \cite{Zhu_co2_red}.  Metal nanoparticles are key components of modern materials for sensing, and the edges of nanoparticles play important role on the performance of the sensors \cite{clement15,hernandez18}. Nanoparticle edges, and in particular edges of Au nanoparticles, are important in two interconnected major problems of our times, namely the quality of air in urban environments through catalytic conversion and sensing of toxic gases, and CO$_2$ emissions and the greenhouse effect. Therefore, knowledge of edge properties is important for applications that have an impact on everyday life.

As nanoparticles become smaller, the ratio of edge- to surface- and bulk atoms increases. As a result, the effect of edges becomes more prominent, and could affect the shape that the nanoparticle will assume. 
Several theoretical studies have predicted nanoparticle shapes from first-principles calculations, typically within the framework of the Wulff construction \cite{Ringe2011,Barmparis2012, Barmparis2015}. In this framework particle shape is determined by the ratios between surface energies of different $(hkl)$ crystal surfaces, while edge and vertex energies are neglected. Moreover, in contrast to the well-established concept of surface energy, little is known about the edge energy of crystalline solids. 

Even though the importance of the edges in the nanoscale is widely accepted, there is no consensus in the literature yet as to the extent of their effect \cite{Alpay2015,Cao2016}. Theoretical investigations have produced different results for edge energies, the main reason being the different definitions of edge length for nanoparticles, while there are very few experimental data for edge energies. At the same time, many important problems of modern nanotechnology depend on the properties and energetics of nanoparticle edges. For example, the importance of the edge and vertex atoms has been observed experimentally in the work of Campbell \textit{et al.}\cite{Campbell2002}, where they found that their calorimetric data on Ru particle formation did not match a model that took into account only the surface energy, especially for smaller nanoparticles. Additionally, Alpay \textit{et al.}\cite{Alpay2015} showcased electronic microscopy images of nanoparticles where their corners are rounded, proving that vertex atoms have considerably large energy and therefore are avoided in the equilibrium shape.

Edge energy is usually defined as the energy required to form an edge divided by the edge length, in the limit of infinite edge length. To compute edge energy, one needs also the surface energy as any partition of an infinite bulk crystal generates both surfaces and edges. Different researchers have used different definitions and methods to calculate the edge energy of a given crystal at a given orientation. Hamilton \cite{Hamilton2006} tried 3 different definitions for edge energy: (i) taking all the atoms in the edge leading to $L = n d$ ($d$ is the distance between first neighbours), (ii) not include half of the two atoms at the vertices of the edge with $L = (n-1) d$ and (iii) a compromise between the two with $L = (n-0.5)d$.  Using Molecular Dynamics (MD) calculations with semi-empirical potentials on Pd clusters, he arrived at values in the order of some meV\AA$^{-1}$, which was an order of magnitude lower than the value expected based on the calculated surface energies. Pelaez \textit{et al.} \cite{Pelaez2012} investigated 
the edge energy of Ni and Al using nanowires with different facets exposed and MD calculations with Embedded Atom Method (EAM) potentials. They found the edge energy of these structures in the order of 0.1 eV\AA $^{-1}$. \cite{Pelaez2012}. Zhao \textit{et al.}\cite{Zhao2016}, after showcasing the problem of the definition of an edge, introduced the idea of relative edge energy (REE) as the difference in energy over the difference of total edge length of two different structures with the same number of atoms since they attributed most of the energy difference on the edge energy. Using their method on Ru nanoparticles they arrived at values of around 50 meV\AA $^{-1}$ for the edge energy. Lai \textit{et al.}\cite{Lai2020} investigated the dependence of the fractional area of (111) surfaces to the edge energy on truncated octahedra and truncated cube nanostructures, which  only have (111) and (100) surfaces; they used Density-Functional Theory (DFT) for the calculation of surface and edge energy of four different transition metals. They found values for the edge energy ranging from 0.17 to 0.32 eV\AA$^{-1}$.

These pioneering works on edge energy of metals demonstrate that calculation of edge energies is a highly non-trivial task. Different definitions of edge energy and different computational methods result in values that span two orders of magnitude, from a few meV to hundreds of meV. The difficulty in the calculation of the edge energy lies in the definition of the edge length, which is not uniquely defined in nanostructures. The only unambiguous definition of edge energy comes for a nanowire of large enough diameter and infinite length. Here, we propose a method to tackle this challenge by calculating energy per edge atom, which then can be converted to energy per length when the length per atom for an infinite nanowire is known. To get this value, we consider several different nanostructures, with periodicity along two, one or zero directions, and which contain the same types of edge atoms. We use a supervised Machine Learning (ML) approach based on a multiple linear regression algorithm to obtain energies of various types of atoms in nanostructures. The method is verified to give precise results for bulk and surface energies for nine different energy functionals. Some of the functionals perform better than others for the surface energy but since the quantity we are more interested in is the edge energy density, a quantity that has not been measured experimentally yet, it is not expected that the same functional will give the most accurate edge energy density. Hence, here we present values calculated for this quantity with a variety of functionals that can be compared when an experimental estimation of this elusive quantity is achieved in the future. From the energies calculated from the presented model the edge energy density and vertex energies of the studied geometries can be extracted. From the energies calculated from the presented model, the edge energy density and vertex energies of the studied geometries can be extracted. The values are well converged with respect to system size.

\section{Methodology}
We focus on Au nanoparticles, as Au is among the most commonly used metals in modern nanotechnology, and the properties we calculate here may be useful for potential applications. Moreover, many different interatomic potentials exist for Au, and this allows for extensive testing of the present model. However, the methodology presented here can be easily transferred to any other metal.

We begin by constructing databases of atomic configurations and their total energies  that will be used with the Machine Learning algorithm. To this end, we calculate the total energy of many different nanostructures. The total energy is calculated either by quantum-mechanical simulations at the level of Density-Functional Theory (DFT), or by classical simulations using a variety of interatomic potentials typically used in Molecular Dynamics (MD) simulations. 

DFT calculations were performed with the the Vienna ab-initio simulation package (VASP) \cite{Kresse1993,Kresse1999, Kresse1996,Kresse1996a}, using the Projector Augmented Wave (PAW)\cite{Blochl1994} method (potentials version of Sep. 2000) and the Generalized Gradient Approximation of Perdew-Burke-Ernzerhof (PBE) \cite{Perdew1996} for the exchange-correlation functional. 
A 500 eV cut-off energy was used for all the calculations. A single ${\mathbf k}$-point was used to sample the Brilluin zone of the nanoparticles while for the periodic structures of nanowires and surface slabs a k-point sampling of ($15\times1\times1$) and ($15\times15\times1$) was used
respectively. All atomic coordinates were allowed to relax to reach the minimum total energy of the system. 
For all DFT calculations, the lattice constant used was the one we found from DFT bulk relaxation, which was found to be 4.173 \AA, close to the experimental lattice constant of Au which is 4.08 \AA. 

MD calculations were performed using the Large-scale Atomic/Molecular Massively Parallel Simulator (LAMMPS) package. For these calculations the experimental lattice constant of 4.08 \AA{} was used. We use eight different interatomic potentials for Au: The simple pair potentials of Lennard-Jones \cite{Erkoc1997} and Morse \cite{Erkoc1997}, the many-body Effective Medium Theory (EMT) potential of Jacobsen\cite{Jacobsen1996} and five different many-body Embedded-Atom Method (EAM) potentials\cite{Ackland1987, Zhou2004, Foiles1986, Olsson2010, Grochola2005}.  

\begin{figure}
	
	\label{structures}
	\includegraphics[scale = 0.20]{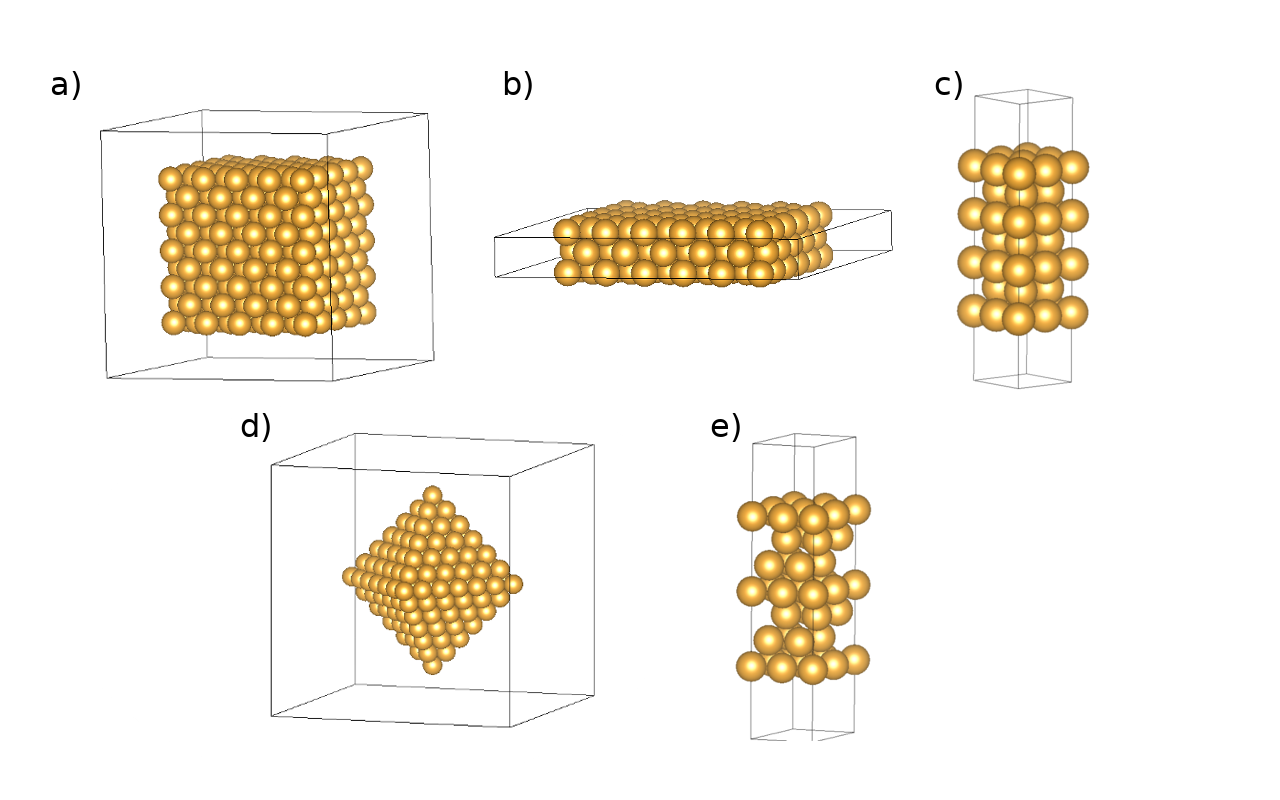}
	\caption{Typical samples from the 5 types of structures that were used to produced data for the ML algorithm. The computational cell is also shown. a) cubic ((100) exposed facets) nanoparticle, b) nanowire with the (100) facets exposed, extending infinitely vertically, c) (100) surface slab extending infinitely in the two lateral axes. d) octahedral ((111) exposed facets) nanoparticle and e) (111) surface slab extending infinitely in the two lateral axes. Figure was created using the VESTA software \cite{vesta}.}
	\label{fig:structures}
\end{figure}

Databases were constructed for the ML algorithm using DFT and MD calculations of the total energy for various nanostructures, such as slabs, nanowires and high-symmetry nanoparticles.
We consider two different classes of nanostructures. The first class comprises systems that only contain (100) faces and edges between these (100) faces; these structures are used to extract the edge energy of the (100)/(100) edge and the vertex energy of the (100)/(100)/(100) vertex. This class contains slabs with surfaces parallel to the (100) plane, tetragonal nanowires of infinite length with surfaces parallel to (100) and the equivalent (010) planes, and cubic nanoparticles with faces parallel to (100) and the equivalent (010) and (001) planes. In fcc Au, as well as any metal with cubic point group symmetry, the (100), (010) and (001) planes have identical atomic structure. These structures are shown in Fig. \ref{fig:structures}. The second class of nanostructures contains systems that contain only (111) faces and edges between these (111) faces; these structures are used to extract the edge energy of the (111)/(111) edge and the vertex energy of the (111)/(111)/(111) vertex.

\begin{table}
    \centering
    \caption{Summary of the nanostructures contained in each database. The MD potential rows represent all eight of MD potentials used for both the relaxed and unrelaxed structures.}
    \begin{tabular}{ccc}
         \hline 
         Exposed Facet   & Nanostructures & Datapoints \\ \hline \hline
         (100) DFT unrelaxed & nanoparticles, nanowires, slabs &  22 \\   
         (100) DFT relaxed   & nanoparticles, nanowires, slabs &  15 \\ 
         (100) MD potential  & nanoparticles                   &  28 \\
         (111) DFT unrelaxed & nanoparticles, slabs            &  13 \\ 
         (111) DFT relaxed   & nanoparticles, slabs            &  9  \\
         (111) MD potential  & nanoparticles                   &  48 \\ \hline
         
    \end{tabular}
    
    \label{tab:db_summary}
\end{table}
This class contains slabs with surfaces parallel to the (111) plane and octahedral nanoparticles  with faces parallel to (111) and equivalent planes. In fcc Au, as well as any metal with cubic point group symmetry, the eight (hkl) planes with $h,k,l = \pm 1$ have identical atomic structure to the (111) plane. We consider regular octahedra with eight faces parallel to these eight planes, twelve  edges and six vertices each.

DFT calculations are limited to few hundreds of atoms due to very high computational cost, while MD can easily handle several millions of atoms. Due to the vast number of entries in each MD database, it was not necessary to include results from slabs and nanowires, and these structures are used only in DFT calculations. In total, we consider several tens of different systems for DFT calculations and about 70 nanoparticles for each classical potential calculations. 

In total, we constructed about seven hundred different data-points divided in 36 datasets corresponding to different structures or different calculation methods. 
Once the databases are made, we extract features of each structure and use machine learning (ML) with the target property being the total energy. The linear regression algorithm was used, as is implemented in the Python Scikit-Learn package (version 1.0.2)\cite{scikit-learn}. Multiple linear regression uses the formula of Eq. (\ref{linear_regression}).
\begin{equation}
    \label{linear_regression}
    y = a_0 + a_1x_1 + a_2x_2 + ... + a_Nx_N.
\end{equation}
Here y is the target property, N the number of features and $x_1$, $x_2$, ..., $x_N$  are the features/independent variables. Multiple Linear Regression fits the coefficients of the model $a_0$, $a_1$, $a_2$, ..., $a_N$ to minimize the residual sum of squares between the predicted y and the real y. For scoring the $R^2$ metric was used which is given by:
\begin{equation}
    \label{r_sq}
    R^2 = 1 - \frac{\sum(y_{pred}-y_{real})^2}{\sum(y_{real}-\bar{y}_{real})^2},
\end{equation}
in which the sums are over all the target values and $\bar{y}_{real}$ is defined as the mean value of the real values:

\begin{equation}
    \bar{y}_{real} = \frac{\sum(y_{real_i})}{N_{real}}
\end{equation}

\section{Results and Discussion}

\subsection{Model for total energy as a function of shape}

We advocate that the total energy of each nanostructure equals the sum of bulk, surface, edge, and vertex energies. While the decomposition of the energy into bulk and surface energy is well established for materials since the nineteenth century \cite{rosakis14,Barmparis2015,rosakis16}, a further decomposition of surface energy into contributions from planar surfaces, edges and vertices is still under discussion.  
\newline \newline
We start from the Gibbs free energy of a structure that can be expressed as 
\begin{equation}
    \label{Gibbs_energy}
    E = E_{bulk} + \sum_{hkl} \gamma_{hkl}A_{hkl},
\end{equation}
in which $\gamma_{hkl}$ and $A_{hkl}$ are the surface tension and surface area of the surface with (hkl) Miller indices. In the case of the structures studied in the present work only one type of (hkl) surfaces are exposed and hence Eq. \ref{Gibbs_energy} can be rewritten as
\begin{equation}
    \label{Gibbs_2}
    E = E_{bulk} + \gamma A ,
\end{equation}
where $\gamma$ is the surface tension of the exposed facets and A is the total surface area. In Eq. (\ref{Gibbs_energy}) and Eq. (\ref{Gibbs_2}) the contributions of the edges and vertices can be considered to be included in the surface term. If these contributions are decomposed from the surface term, the Gibbs equation takes the form
\begin{equation}
    \label{Gibbs_3}
    E = E_{bulk} + \gamma A' + \tau L + N_v v .
\end{equation}
Here, $A'$ is the surface without counting the edge and vertex atoms, $\tau$ is the edge energy per unit length, $L$ the total edge length, $N_v$ is the number of vertices and $v$ is the energy of a vertex atom. Again in the structures studied, only one type of edge and vertices is present but even if this was not the case the idea can be generalized with a sum like Eq. (\ref{Gibbs_energy}).

In order to use Eq. (\ref{Gibbs_3}) for the calculation of  $E_{bulk}, \gamma, \tau$ and $v$, one has to perform many different calculations for the total energy, $E$, of different shapes and sizes of nanostructures, and then do some sort of fitting of the results. The obvious choice is to try to transform Eq. (\ref{Gibbs_3}) into a polynomial form with only one variable. We tried fitting the energies to a polynomial of $x=N^{1/3}$ where $N$ is the total number of atoms. For large $x$, Eq. (\ref{Gibbs_3}) would be a cubic polynomial of $x$ as $L\sim x, A\sim x^2$ and $V\sim x^3$. Zhao \textit{et al.}\cite{Zhao2016} used a similar idea by expressing all terms of Eq.  (\ref{Gibbs_3}) with respect to the total edge length $L$. In both their case and in the present work, the fit of the simulation data to the equation yielded unsatisfactory results, which in many cases also had the wrong signs e.g. negative vertex or edge energy. Additionally, since the equation was polynomial in nature, the extracted coefficients depended strongly on the initial guess and even after imposing constraints based on physical intuition the results were still not promising.

Considering the above limitations, a different approach was adopted: assuming that an energy decomposition exists, we can express the total energy as a linear function of the numbers of atoms of bulk, planar surfaces, edges, and vertices. 
In the following, we will refer to  atoms on planar surfaces, simply as "surface atoms", explicitly excluding edge and vertex atoms. 

The atoms of the nanostructure are characterized as bulk, surface 
edge, and vertex atoms based on their position with respect to the symmetry elements of each nanostructure, as well as their coordination number, $z$. The later is equal to the number of first neighbours of the given atom. The coordination number $z$, is maximum for bulk atoms and becomes lower as we consider surface, edge and vertex atoms, respectively.  For fcc Au, these numbers range from $z=12$ for bulk atoms down to $z=3$ for atoms at the vertices of cubic nanoparticles. 

We assume that the total energy of the nanoparticle is linear with respect to the four numbers of atoms, that is,
\begin{equation}
	\label{energy_decomp_equation}
	E = E_b N_b + E_s N_s + E_e N_e + E_v N_v,
\end{equation}
while the total number of atoms in the nanoparticle, $N$, is the sum of the numbers of the four different kinds of atoms:
\begin{equation}
	\label{number_of_atoms}
    N = N_b + N_s + N_e + N_v.
\end{equation}

Equation (\ref{energy_decomp_equation}) is a special case of a more general decomposition by considering all coordination numbers, $z$: 

\begin{equation}
    E = \sum_z N_z E_z
    \label{general_decomp_equation}
\end{equation}

 where $N_z, E_z$ are number and energy of atoms with coordination number $z$, an approach that has been proved to work extremely well in a variety of cases recently: Holec \textit{et al.}\cite{Holec2020} used this differentiation between the atoms to find the average bond energy. Vega \textit{et al.}\cite{Vega2021} performed similar decomposition in terms of the fractions of the uncoordinated atoms to the bulk for Pd nanoparticles. Roling et al. used a robust energy decomposition scheme to predict the stability of nanoparticles that span a wide-ranging combinatorial space \cite{roling18, Roling19}. Here, we use the relatively simpler Eq. (\ref{energy_decomp_equation}) which has the advantage that it provides a straightforward connection between atomic energies and continuum properties, including bulk-, surface-, edge- and vertex energies.

Similar decompositions of the energy into various types of atoms have been used in the past for other problems, for example for the thermodynamic description of  size dependent shape evolution \cite{Barnard2004}, for the explanation of enhanced catalytic activity of nanoparticles \cite{lopez04}, or to account for the mechanical properties of nanocrystalline solids \cite{galanis13}. Eq.  (\ref{energy_decomp_equation}) should give satisfactory predictions if the interatomic interactions were described by pair potentials, and is expected to hold very well for noble metals, like Au, whose atoms possess filled $d$ shells and are therefore spherically symmetric. Indeed, we find that Eq.  (\ref{energy_decomp_equation}) is valid for Au for all cases considered in the present study.

\subsection{Atomistic simulations}

\begin{figure}
	\includegraphics[width=\columnwidth]{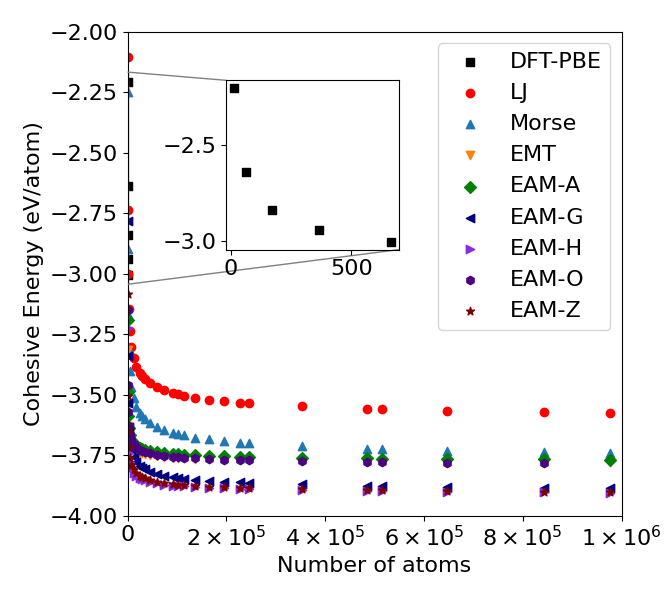}
	\includegraphics[width=\columnwidth]{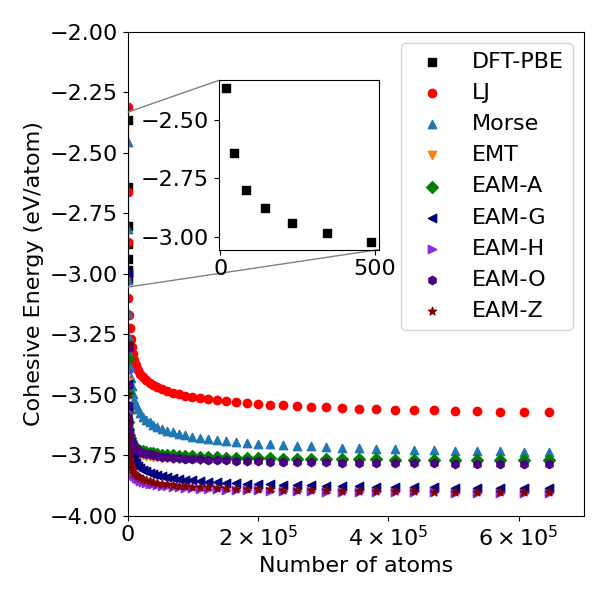}
	\caption{Cohesive Energy of the unrelaxed structures as calculated with the various methods with respect to the total number of atoms for Up: cubic nanoparticles and Down: octahedral nanoparticles. 
	}
		\label{Cohesive_energy}
\end{figure}

We start by calculating the total energy of all structures, using both MD and DFT codes. We then repeat the calculation allowing the atomic positions to relax and minimize the total energy of the system. As expected, a relatively small percentage of atoms, mostly at the outmost layers change their positions during relaxation, and the overall shape is well preserved in the process. For each structure, we calculate the cohesive energy, defined as 
\begin{equation}
	\label{cohesive energy}
	E_c = E/N, 
\end{equation}
where $N$ is the total number of atoms and $E$ is the calculated total energy. All energies are reported with reference to the energy of isolated atom: For all calculation methods, the energy of one atom at the center of a very large simulation box is zero.
Fig.  \ref{Cohesive_energy} shows the results for the cohesive energy for cubic and octahedral nanoparticles.

For a given nanoparticle, the number of vertex atoms is always the same, independent of size, while the ratio of edge to bulk and surface to bulk declines as the total number of atoms increases. For large nanoparticles of average diameter $d$, the number of atoms, $N$ is proportional to $d^3$ as the particle density, $N/V$ is constant and the volume, $V$ is proportional to $d^3$. For similar reasons, $N_s$ will be proportional to $d^2$ and $N_e$ will be proportional to $d$. Therefore, for a large nanoparticle with $N$ atoms, $N_b, N_s, N_e$ and $N_v$ will be proportional to $N, N^{2/3}, N^{1/3}$ and $N^0$, respectively. 

For this reason, the cohesive energy reaches a constant value for large values of $N$, which is independent of shape and is equal to the bulk cohesive energy, which in turn should equal the coefficient $E_b$ of Eq.  (\ref{energy_decomp_equation}). As the value of $E_b$ is model-dependent, the various methods give slightly different asymptotic value for the cohesive energy, as shown in Fig.  \ref{Cohesive_energy}.

\begin{figure}
	\includegraphics[scale=0.42]{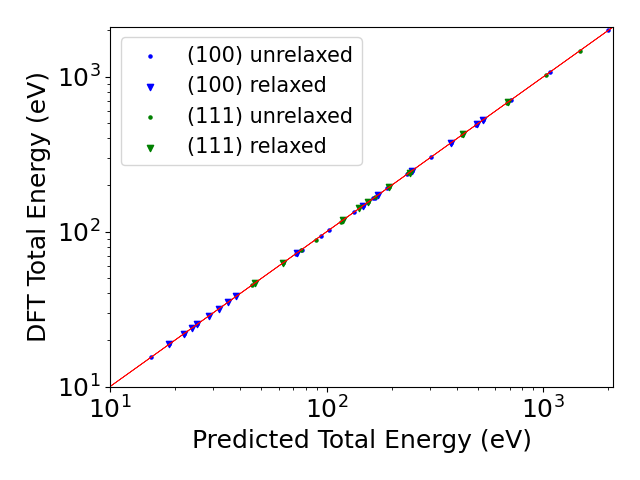}
	\includegraphics[scale=0.42]{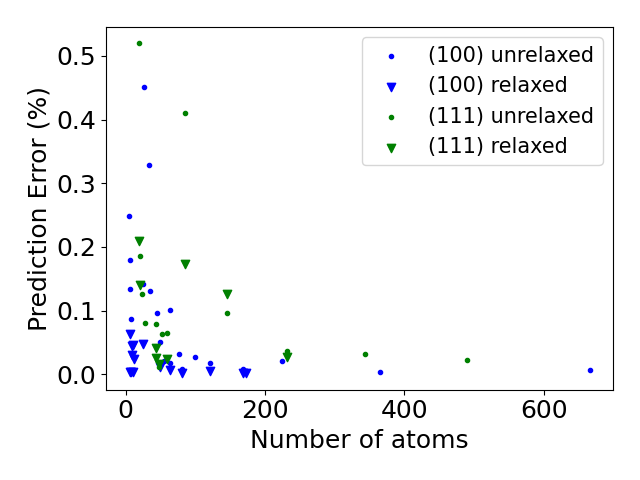}
	\caption{Up: The calculated DFT total energy versus predicted total energy for the DFT datasets. Down: The percentage error versus the number of atoms.}
	\label{ml_dft}
\end{figure}

\subsection{Machine learning computations}

Databases were constructed for DFT and each different MD interatomic potential, using the number of each atom type and the total energy of each nanostructure. For the ML model, the features are the number of each atom type ($N_b$, $N_s$, $N_e$, $N_v$) and the property is the total energy of the nanoparticle. In the linear model, the coefficients of the fit are the various energies $E_b, E_s, E_e, E_v$ of Eq.  (\ref{energy_decomp_equation}). 

The ML model that used Eq.(\ref{energy_decomp_equation}) to fit the data produced an excellent score, $R^2>0.99$, for all the datasets. Interestingly enough, the relatively simple energy decomposition model was quite accurate even in cases that wasn't expected, like the DFT based datasets, both unrelaxed and relaxed, in which the model reproduced the total energy with a percentage error of less than  0.5\%. The results can be seen in Fig. \ref{ml_dft}. For Fig. \ref{ml_dft}, all the datapoints were used to extract the energy coefficients of Eq. (\ref{energy_decomp_equation}) and then used to predict the energy of all the nanostructures. On the right of Fig. \ref{ml_dft} the percentage error of the prediction is plotted versus the size of the nanostructures in which a relatively small size effect is revealed with the smaller nanostructures giving rise to higher deviations in the prediction but the prediction accuracy is still excellent. 

The ML-predicted values for the $E_b$, $E_s$, $E_e$ and $E_v$ of all the MD potentials were quite robust and did not change with varying the dataset size. Additionally, as mentioned in section A we can also use the generalized Eq. (\ref{general_decomp_equation}) that allows us to add all our datapoints in one database and doing so left the calculated parameters almost unchanged, attesting to the robustness and accuracy of the model. Leave-One-Out cross-validation yielded excellent predictions for the MD datasets with a mean absolute percentage error in the order of $10^{-7}$ for the unrelaxed datasets while for the relaxed ones, the same error was in the order of $10^{-3}$  while the same cross-validation for the DFT datasets yielded a mean absolute percentage error in the order of $10^{-3}$ for both relaxed and unrelaxed structures. 
\begin{table}

	\caption{Average energy of bulk, surface, edge and vertex atoms in unrelaxed and relaxed cubic and octahedral nanostructures. Cubes expose (100) surfaces; octahedra expose (111) surfaces. All energies are given in eV/atom and the value in the parentheses represents the difference percentage to the relevant DFT value.}
	\begin{center}
	
		{\bf (100)-oriented (cubes), unrelaxed}
		
		\begin{tabular}{ccccc}
			\hline
			Potential & $E_b$  & $E_s$ & $E_e$  & $E_v$  \\ \hline \hline
			
			LJ\cite{Erkoc1997} & -3.65 (0.12) & -2.17 (-0.22) & -1.22 (-0.47) & -0.67 (-0.58) \\
			MORSE\cite{Erkoc1997} & -3.81 (0.17) & -2.34 (-0.16) & -1.34 (-0.42) & -0.74 (-0.53) \\
			EMT\cite{Jacobsen1996} & -3.80 (0.16) & -3.45 (0.24) & -3.00 (0.30) & -2.50 (0.57)\\
			EAM-A\cite{Ackland1987} & -3.79 (0.16) & -3.38 (0.22) & -2.88 (0.25) & -2.01 (0.26)\\
			EAM-Z\cite{Zhou2004} & -3.93 (0.20) & -3.37(0.21) & -2.52 (0.09) & -1.50 (-0.06)\\
			EAM-F\cite{Foiles1986} & -3.93 (0.20) & -3.45 (0.24) & -2.74 (0.19) & -2.07 (0.30)\\
			EAM-O\cite{Olsson2010} & -3.81 (0.17) & -3.31 (0.19) & -2.71 (0.17) & -2.12 (0.33)\\
			EAM-G\cite{Grochola2005} & -3.92 (0.20) & -3.16 (0.14) & -1.84 (-0.20) & -0.91 (-0.43)\\
			DFT-PBE & -3.27 (0.00) & -2.78 (0.00) & -2.31 (0.00) & -1.59 (0.00)\\
			
		\end{tabular}
		\vspace{1.0em}
		
		{\bf  (100)-oriented (cubes), relaxed}
		
		\begin{tabular}{ccccc}

			\hline
			Potential & $E_b$  & $E_s$ & $E_e$  & $E_v$  \\ \hline \hline
			LJ\cite{Erkoc1997} & -3.65 (0.12) & -2.17 (-0.24) & -1.22 (-0.48) & -0.66 (-0.68)\\
			MORSE\cite{Erkoc1997} & -3.81 (0.17) & -2.35 (-0.17) & -1.37 (-0.42) & -0.74 (-0.64)\\
			EMT\cite{Jacobsen1996} & -3.80 (0.16) & -3.50 (0.24) & -3.28 (0.39) & -2.83 (0.36)\\
			EAM-A\cite{Ackland1987} & -3.79 (0.16) & -3.39 (0.20) & -3.12 (0.32) & -1.89 (-0.09)\\
			EAM-Z\cite{Zhou2004} & -3.93 (0.20) & -3.41 (0.21) & -2.99 (0.27) & -1.59 (-0.24)\\
			EAM-F\cite{Foiles1986} & -3.93 (0.20) & -3.46 (0.23) & -3.13 (0.33) & -2.36 (0.13)\\
			EAM-O\cite{Olsson2010} & -3.81 (0.17) & -3.33 (0.18) & -3.00 (0.27) & -2.37 (0.14)\\
			EAM-G\cite{Grochola2005} & -3.92 (0.20) & -3.26 (0.16) & -2.48 (0.05) & -0.83 (-0.60)\\
			DFT-PBE & -3.27 (0.00) & -2.82 (0.00) & -2.36 (0.00) & -2.08 (0.00)\\

		\end{tabular}

		\vspace{2.0em}
		{\bf (111)-oriented (octahedra), unrelaxed}
		
		\begin{tabular}{ccccc}
			\hline
			Potential & $E_b$  & $E_s$ & $E_e$  & $E_v$  \\ \hline \hline 
			LJ\cite{Erkoc1997} & -3.65 (0.12) & -2.42 (-0.17) & -1.8 (-0.32) & -0.93 (-0.46)\\
            MORSE\cite{Erkoc1997} & -3.81 (0.17) & -2.58 (-0.11) & -1.89 (-0.28) & -1.00 (-0.42)\\
            EMT\cite{Jacobsen1996} & -3.80 (0.16) & -3.54 (0.22) & -3.31 (0.26) & -2.77 (0.60)\\
            EAM-A\cite{Ackland1987} & -3.79 (0.16) & -3.50 (0.20) & -3.23 (0.23) & -1.94 (0.12)\\
            EAM-Z\cite{Zhou2004} & -3.93 (0.20) & -3.50 (0.20) & -3.06 (0.16) & -1.63 (-0.06)\\
            EAM-F\cite{Foiles1986} & -3.93 (0.20) & -3.56 (0.22) & -3.20 (0.22) & -2.40 (0.39)\\
            EAM-O\cite{Olsson2010} & -3.81 (0.17) & -3.43 (0.18) & -3.10 (0.18) & -2.41 (0.39)\\
            EAM-G\cite{Grochola2005} & -3.92 (0.20) & -3.34 (0.15) & -2.56 (-0.03) & -1.32 (-0.24)\\
            DFT-PBE & -3.27 (0.00) & -2.91 (0.00) & -2.63 (0.00) & -1.73 (0.00)\\
		\end{tabular}
		\vspace{1.0em}
		
		{\bf  (111)-oriented (octahedra), relaxed}
		
		\begin{tabular}{ccccc}
			\hline
			Potential & $E_b$  & $E_s$ & $E_e$  & $E_v$  \\ \hline \hline 
			LJ\cite{Erkoc1997} & -3.65 (0.12) & -2.42 (-0.18) & -1.76 (-0.34) & -0.88 (-0.62)\\
            MORSE\cite{Erkoc1997} & -3.81 (0.17) & -2.58 (-0.13) & -1.93 (-0.28) & -0.89 (-0.61)\\
            EMT\cite{Jacobsen1996} & -3.80 (0.16) & -3.57 (0.21) & -3.44 (0.29) & -2.97 (0.30)\\
            EAM-A\cite{Ackland1987} & -3.79 (0.16) & -3.51 (0.19) & -3.3 (0.24) & -2.79 (0.22)\\
            EAM-Z\cite{Zhou2004} & -3.93 (0.20) & -3.52 (0.19) & -3.28 (0.23) & -2.06 (0.10)\\
            EAM-F\cite{Foiles1986} & -3.93 (0.20) & -3.58 (0.21) & -3.38 (0.27) & -2.48 (0.08)\\
            EAM-O\cite{Olsson2010} & -3.81 (0.17) & -3.44 (0.17) & -3.24 (0.21) & -2.49 (0.09)\\
            EAM-G\cite{Grochola2005} & -3.92 (0.20) & -3.39 (0.15) & -3.00 (0.12) & -0.30 (-0.87)\\
            DFT-PBE &                  -3.27 (0.00) & -2.95 (0.00) & -2.67 (0.00) & -2.29 (0.00)\\
		\end{tabular}
	\end{center}
	\label{tab:ene}
\end{table}

\subsection{Linear regression}

We consider four different datasets: relaxed/unrelaxed and (100)-oriented/(111)-oriented structures. The calculated coefficients for all datasets are summarized in Table \ref{tab:ene}. In all cases, the relation between the coefficients was 
\begin{equation}
E_b < E_s < E_e < E_v < 0
\label{eq_inequality}
\end{equation}
as expected, since the number of bonds for the atom types is in decreasing order: bulk, surface, edge, vertex. To a first approximation, energy is lower for larger number of bonds, therefore energies of Au atoms should increase with decreasing coordination number. Also, in all cases energies of Au atoms in a relaxed structure should be lower than energies of isolated Au atoms which is set to zero in this study. Preserving the ordering of Eq.  (\ref{eq_inequality}) is not trivial; in many cases, numerical errors cause some coefficients to have wrong ordering or even be positive \cite{Zhao2016}.

In all cases, the bulk energy, $E_b$ is the same in all four tables and changes by less than 10 meV/atom upon relaxation, as the relaxation does not affect very much atoms with bulk coordination. 
The experimental value of the cohesive energy of Au is -3.81 eV/atom \cite{Kittel2005}. Most empirical potentials reproduce this value or give values that are very close to it. The DFT value of  -3.27 eV/atom deviates from the experimental value by -14\%, in accordance to general trends in DFT calculations for the cohesive energy of metals \cite{Tran2016}.

For the surface energy, $E_s$, the ML regression predicts correctly that it is lower for (111) surfaces found in octahedra compared to (100) surfaces found in cubes; this holds for all potentials. Moreover, in all cases the relaxed surface energy is lower than the unrelaxed surface energy, as expected. 

The experimental value for the surface tension of Au(111) is $\gamma=1.5$ J/m$^2$ \cite{Tyson1977}. In order to translate this value to the notation used in the present work, we use the definition that surface tension is the excess energy related to bulk divided by the area, $A$, i.e.,
\begin{equation}
	\gamma = \frac{E - NE_b}{A} \quad \Rightarrow \quad \gamma =  \frac{E_s-E_b}{A_{at}}.
	\label{eq:gamma} 
\end{equation}
In Eq. (\ref{eq:gamma}), the area per atom, $A_{at}=A/N_s$ for Au(111) equals $A_{at}=a^2\sqrt{3}/4$ where $a=4.08$ \AA\ is the experimental lattice constant of Au. The second equation in  \ref{eq:gamma} holds for slabs, where $N_e=N_v=0$, or in the limit of large nanoparticles,  where $N_e$ and $N_v$ are negligible.

Using experimental values for the cohesive energy ($E_b$), surface tension ($\gamma$) and $A_{at}$ for (111), we obtain the experimental value of $E_s$ to be $E_s = -3.13$ eV/atom. Contrary to the case with the cohesive energy, the empirical potentials, that are fitted to experimental values, give slightly worse result for the surface energy compared to the first-principles DFT calculation which gives $E_s=-2.95$ eV. Notice that the DFT values for $E_s$ is higher than the experimental value; however, as the DFT bulk energy, $E_b=-3.27$ eV, is also higher than the experimental cohesive energy of Au, $E_b=-3.81$ eV, the DFT value for surface tension, $\gamma$, of Au(111) turns out to be lower than the experimental value and deviates more than the empirical potentials, so in summary due to error cancellation the DFT value of $E_s$ is closer to experiment than by using $\gamma$. The values of surface tensions for Au(111) and Au(100) calculated from the DFT values for $E_s$, are 0.68 J/m$^2$ and 0.83 J/m$^2$ respectively, in excellent agreement to previously reported DFT values from slab calculations \cite{Barmparis2012}.

The vertex energy, $E_v$, is quite sensitive to numerical errors: the nanoparticles used in the empirical potential calculations of the present study contain many millions of atoms and only eight vertex atoms  (cubes) or six vertex atoms (octahedra).
Therefore, vertex energy is a quantity that is extremely vulnerable to numerical errors in empirical potential calculations. 
EMT and some EAM potentials give reasonable vertex energies that are close to the DFT values (Table \ref{tab:ene}). The relative vertex energies $v=E_v-E_b$ from the DFT calculations are 
\begin{center}
	$v_{(100)/(100)/(100)}$ = 1.19 eV,
\end{center}
\begin{center}
	$v_{(111)/(111)/(111)}$ = 0.98 eV
\end{center}

 As expected, the vertex formed by three (111) surfaces, where the Au atom has four neighbors has lower energy than the vertex formed by three (100) surfaces, where the Au atom has three neighbors. It should be noted that, for the case of $v_{(111)/(111)/(111)}$, including smaller nanoparticles with less than 90 atoms in the database causes deviation of up to 0.5 eV from the reported value. The reason behind this deviation probably stems from the fact that in these nanoparticles there are few bulk atoms that are also relatively close to the surfaces due to the acute angles of the octahedra. On the contrary, all other values  for energies of bulk, surface, edge atoms, and the energy of the (100)/(100)/(100) vertex were found to be well-converged within the present dataset, and any modifications of the dataset size result in changes of few meV per atom at most.

\begin{table}
	\caption{Edge energies of the (100)/(100) and (111)/(111) edges of Au, in eV/\AA, as calculated by machine-learning linear-regression algorithm based on DFT-PBE data. Values for relaxed and unrelaxed structures are given. }
	\begin{center}
		\begin{tabular}{ccccc}
			\hline
			Edge type & unrelaxed & relaxed \\ \hline \hline
			(100)/(100) & 0.23 & 0.22  \\
			(111)/(111) & 0.22 & 0.20  \\
		\end{tabular}
	\end{center}
	\label{tab:tau}
\end{table}

\subsection{Edge energies of Au}

The trends for edge energy are similar to the trends observed for surface energy and the values of Table \ref{tab:ene} are ranked as (cube, unrelaxed) $>$ (cube, relaxed) $>$ (octahedron, unrelaxed) $>$ (octahedron, relaxed). For the DFT results, atoms at the edges between (111) surfaces have lower energy by about 10\% lower than the energy of edges between (100) surfaces. 

The edge energy density, $\tau$, is defined as the excess energy over bulk- and surface energy divided by the total length of edges, $L$. For a system that contains edges, Eq.  (\ref{eq:gamma}) is generalized to
\begin{equation}
	\tau = 	\frac{E - NE_b-N_s(E_s-E_b)}{L}. 
	\label{eq:tau} 
\end{equation}
In Eq. (\ref{eq:tau}), the term $E_s-E_b$ is the energy difference between a surface and a bulk atom and is equal to the $\gamma A_{at}$ term of Eq. (\ref{eq:gamma}). Using the decomposition for the total energy of Eq. (\ref{energy_decomp_equation}) and the total number of atoms from Eq. (\ref{number_of_atoms}), one obtains: 

\begin{equation}
	\tau = 	\frac{E - NE_b-N_s(E_s-E_b)}{L} \quad \Rightarrow \quad \tau =  \frac{E_e-E_b}{D_{at}}.
	\label{eq:tau2} 
\end{equation}

The second equation above holds for nanowires, where $N_v=0$, or in the limit of large nanoparticles, where $N_v$ is negligible. $D_{at} = L/N_e$ represents the distance between neighboring edge atoms. For (100)-oriented nanowires $D_{at} = a = 4.173$ while for (111)/(111) edges it is $D_{at} = a\sqrt{2}/2 = 2.951$ \AA. Using this value and the DFT values for $E_e$ and $E_b$, we can extract the edge energy densities, $\tau$, of Au shown in Table \ref{tab:tau}. As expected, the close-packed (111) surfaces that have lower surface energy form edges that are energetically favoured compared to (100). The relaxed values are $\tau_{(100)/(100)} = 0.22$ eV/\AA\ and $\tau_{(111)/(111)} = 0.20$ eV/\AA. As was the case with surface tension, $\gamma$, edge energy density is affected very little by atomic relaxation. It is noteworthy that even though there is a 0.3 eV/atom difference in the values of $E_e$ of cubes and octahedra, the edge energy density is very similar for these edges owning to the more dense octahedra edges. 

The calculated edge energies are in agreement with other published DFT values. Holec \textit{et al.}\cite{Holec2017} used the extended surfaces concept, which should introduce a small size-dependency on the results; applying their idea to the smallest nanoparticle in our dataset, we obtain 0.18eV/\AA\ for the edge energy of Au$_{63}$, close to 0.22 eV/\AA\ of Table \ref{tab:tau} while the difference diminishes for larger nanoparticles. The same value of 0.18 eV/\AA\ is obtained by Lai \textit{et al.}\cite{Lai2020} using DFT-PBEsol calculations on Au nanowires and nanoparticles. Vega \textit{et al.}\cite{Vega2021} report 0.27 eV/\AA\ for the (100)/(100) edge energy of Pd. Holec \textit{et al.}\cite{Holec2020} used the MD potential of Grochola \cite{Grochola2005} and calculated the excess energies, $E_{excess} = E_z - E_b$, of atoms as a function of coordination number, $z$. When we use the same potential, we find an identical result of 0.66 eV for surface atoms with $z=8$ and very similar results, 1.44 eV vs 1.50 eV, for edge atoms with $z=5$. Roling \textit{et al.}\cite{Roling2017} did a similar calculation for Au, using DFT-rPBE and found $E_5$ = 2.05 eV for $z=5$ and 2.21 eV for $z=7$, corresponding to cube and octahedra edges respectively, which compare very well to values of -2.36 eV and -2.67 eV, calculated with PBE, respectively as shown in Table \ref{tab:ene}. In all the above mentioned works in the literature, even though the methods used may differ it seems that an agreement is being reached for the edge energies of nanoparticles. 

\section{Conclusions}

In this work we consider decomposing the energy of a gold nanoparticle into contributions from the bulk, surfaces, edges and vertices. We find that such a decomposition is accurate for a variety of systems and many different calculation methods for the total energy. The parameters of the decomposition formula are calculated using machine learning techniques, with the total energy of a Au nanostructure and number of atoms in the bulk, surfaces, edges, and vertices as input. Our model and method are found to be valid for gold nanostructures to a great accuracy and hold not only for pair potentials, but also for complicated many-body potentials, and for DFT. Values for the energies of different atom types are reported for 8 different MD potentials and DFT-GGA. The energies obtained have an excellent accuracy with a mean absolute percentage error of 10$^{-3}$ at worst, far outperforming more conventional approaches of fitting a polynomial uni-variate equation of energy.
 
By fitting calculated total energies of Au nanostructures, we obtain values for cohesive energy, surface tension, edge energy density and vertex energy of Au. These quantities are known to play a pivotal role in the determination of the shape of a nanoparticle,  though the contribution of edge energy density and vertex energies become significant for smaller nanoparticles.

Values for cohesive energy and surface tension presented here agree with other published works. Edge energies lie between 0.20 and 0.23 eV/\AA~depending on the edge type, in agreement with respect to the order of magnitude with some recent works, while vertex energies are of the order of 1 eV. The validity of our edge and vertex energy calculation is further re-enforced by the very accurate total energy calculations that use these quantities.

The present method can be easily generalized to other metals and arbitrary shapes or nanostructures, provided more structures are included in the databases that are used as input to the ML code. The extension of the present method to surfaces other than (100) and (111), that are used as an example here, is straightforward. Modern computational materials science codes allow for easy generation of databases of structures and first-principles total energies that can be used as input to the ML regression. In a future implementation, adsorption on metal surfaces could also be taken into account. As such, the present method could contribute towards the design of nanoparticles with tailored chemical and physical properties.
\newline
\newline
\textbf{\large Funding} 
\newline
\newline
This work was funded by the Hellenic Foundation for Research and Innovation through project MULTIGOLD, grant HFRI-FM17-1303 /  KA10480.
\newline
\newline
\newline
\newline
\textbf{\large Aknowledgements} 
\newline
\newline
We acknowledge computational time granted 	from the National Infrastructures for Research and Technology S.A. (GRNET S.A.) in the National HPC facility, ARIS, under
projects pr007027-NANOGOLD and pr009029-NANO-COMPDESIGN.

\bibliographystyle{RS} 
\bibliography{main}

\end{document}